\RequirePackage{lineno}
\documentclass[aps,prl,twocolumn,superscriptaddress]{revtex4}

\usepackage{graphicx}
\usepackage{bm}
\usepackage{amsmath}
\usepackage{amssymb}
\usepackage{hyperref}

\begin{document}
\title{Selective Conduction of Organic Molecules via Free-Standing Graphene}

\author{Zhao Wang}
\email{zw@gxu.edu.cn}

\affiliation{Department of Physics, Guangxi University, Nanning 530004, China}

\begin{abstract}
A race is held between ten species of organic gas molecules on a graphene substrate driven by thermal gradients via molecular dynamics. Fast conduction of the molecules is observed with selectivity for aromatic compounds. This selectivity stems from the fact that the planar structure of the aromatic molecule helps keep a shorter distance to the substrate, which is the key to the driving force at the gas-solid interface. The drift velocity monotonically increases with decreasing molecule density, with no ballistic transport observable even for a single molecule. A non-linear regime is discovered for the conduction of benzene molecules under large thermal gradients. At low temperature, molecules formed aggregation and move collectively along specific path in the graphene substrate.
\end{abstract}

\maketitle

\section{Introduction}
Directional transport of molecules is a central process for separation of chemicals \cite{Li2015}, drug delivery \cite{Bianco2005}, molecular machines \cite{Erbas-Cakmak2015}, energy conversion and storage \cite{Park2014}, and so forth. Conduction of molecules can be realized via applying temperature gradients along which molecules drift. In liquid or gaseous mixtures, the driving force is known to stem from the difference in kinetic energy between the hot and the cold molecules \cite{Duhr2006}. However, the driving force of thermodiffusion at solid-phase interfaces is under debate. For instance, Barreiro \textit{et al.} reported that motion of a gold cargo attached to a carbon nanotube (CNT) can be actuated by imposing a thermal gradient along the tube \cite{Barreiro2008}. The underlying mechanism of the actuation was reported to be phononic excitations traveling from the hot to the cold region \cite{Guo2013}. Becton and Wang studied the motion of a nanoflake on a graphene sheet and reported that the driving force is a discrepancy in the kinetic energy across the temperature gradient \cite{Becton2014}. Panizon {et al.} measured a driving force independent of the local gradient magnitude on a gold cluster transmitted through scattering with flexural phonon waves in a graphene substrate \cite{Panizon2017}.

Low-dimensional carbons have drawn considerable attention for nanoscale molecule transport thanks to their chemical inertness and peculiar structures \cite{Zhang2019,Sun2016}. Spontaneous water conduction through a CNT channel was first reported in 2001 by Hummer \textit{et al.} using molecular dynamics (MD) simulations \cite{Hummer2001}. Fluid and solid transport through CNTs was experimentally observed in the following years \cite{Regan2004,Holt2006,Sun2006}. Mass transport through CNTs by thermodiffusion was first reported by Schoen \textit{et al.}, who demonstrated via MD that motion of a gold nanoparticle can be induced on CNTs subjected to thermal gradients \cite{Schoen2006}. This was later confirmed by a scanning electron microscopy experiment \cite{Barreiro2008}. MD remains the state-of-the-art method to explore molecular transport at nanoscale numerically \cite{Leng2016,Panizon2017}. 

Most of previous works focus on the transport of solids by low-dimensional carbon nanostructures. In contrast, little is known about the conduction of gas molecules adsorbed on solid nanostructures \cite{Roos2011}. Can organic molecules be conducted on two-dimension (2D) membranes like graphene? If so, what type of molecules will exhibit higher velocity, and where the driving force comes from? To address these questions and in view of the important implications of the transport of organic compounds for a wide range of applications, we simulate thermodiffusion of organic molecules adsorbed on a suspended graphene sheet subjected to a thermal gradient. The first race is run between ten different species of hydrocarbons via MD.

\section{Methods}

\begin{figure}[htp]
\centerline{\includegraphics[width=8cm]{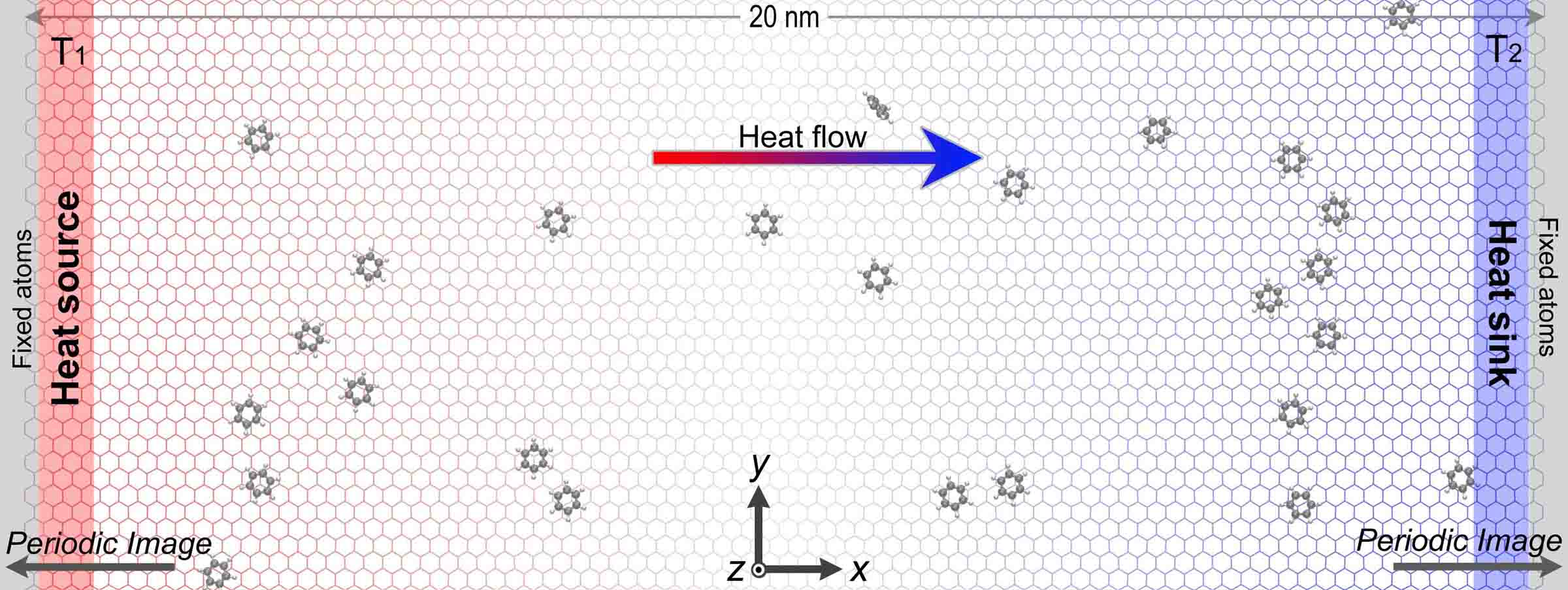}}
\caption{\label{F1}
Schematic of organic molecules adsorbed on a segment of an infinite graphene layer, in which a temperature gradient is applied along the $x$ axis by defining a heat source and a heat sink. This cell is periodic in both the $x$ and $y$ directions.}
\end{figure}

We consider a rectangular cell of $L \times w=20.2 \times 7.8 \;\mathrm{nm}$ in an infinite free-standing graphene layer. In analogy to experiments \cite{Barreiro2008,Zambrano2009,Oyarzua2018}, two thermal energy reservoirs having different temperatures ($T_{1}$ and $T_{2}$) are placed at the both sides of this cell, as shown in Fig.~\ref{F1}. The $x$ and $y$ boundaries of this cell are both set to be periodic in order to ensure a steady flow of molecules. Ten species of small organic molecules (benzene, butane, cyclehexane, cyclopentane, ethane, hexane, methylpropane, pentane, propane and toluene) are chosen to be the adsorbates. Different sets of simulations are performed varying molecular type, adsorbate density, temperature and thermal gradient. In each of these simulations, a number of organic molecules are initially adsorbed at random positions atop the graphene. The atoms of the adsorbates and the free part of graphene are free to move, while the temperatures of the graphene atoms in the thermal reservoir regions are controlled to progressively reach $T_{1}$ and $T_{2}$ by using the Nos\'{e}-Hoover thermostat \cite{Plimpton95,Wang2018,Wu2019,Guo2015,Li2013,Lin2014}. The MD run for a period of $5 \sim 15\;\mathrm{ns}$ with a time step of $0.5\;\mathrm{fs}$ after thermal equilibrium is reached within about $5\;\mathrm{ns}$. The simulation scheme is illustrated in video recordings provided in the Supporting Information.

The potential energy $\varepsilon^{p}$ of the system is given by the adaptive interatomic reactive empirical bond order (AIREBO) potential as a sum of many-body interaction bonds,

\begin{equation}
\label{eq1}
\varepsilon^{p}=\frac{1}{2} \sum\limits_{i=1}^n \sum\limits_{\substack{j=1 \\ j\ne i} }^n 
[ \varepsilon^R_{ij} + b_{ij}\varepsilon^A_{ij} + \varepsilon^{vdw}_{ij} + 
\sum\limits_{\substack{k=1 \\ k\ne i,j} }^n \sum\limits_{\substack{\ell=1 \\ \ell\ne i,j,k}}^n \varepsilon_{kij\ell}^{tor}]
\end{equation}

\noindent where $n$ is the total number of atoms, $i$, $j$, $k$ and $\ell$ are atom index numbers. $\varepsilon^R$ and $\varepsilon^A$ are the interatomic repulsion and attraction terms for the valence electrons, respectively. $\varepsilon^{tor}$ is a single-bond torsion term. Many-body effects are included in the bond-order function,

\begin{equation}
\label{eq3}
b_{ij} = \frac{1}{2} \left( b_{ij}^{\sigma-\pi} + b_{ji}^{\sigma-\pi} + b_{ji}^{RC}+ b_{ji}^{DH} \right)
\end{equation}

\noindent where $b_{ij}^{\sigma-\pi}$ is a function of the atomic distance and bond angle, $b_{ji}^{RC}$ represents effects of the bond conjugation, and $b_{ji}^{DH}$ depends on the dihedral atomic angle. The long-range Van der Waals (vdw) interactions are included by adding the term $\varepsilon^{vdw}$, which is given by a parameterized Lennard-Jones (LJ) force field with a cutoff radius of $1.1\;\mathrm{nm}$. The parameter values and benchmarks of this force field are provided in Ref.\,\citep{Stuart2000}. Compared to other force fields for hydrocarbon systems, the modeling of bond rotation and torsion in the AIREBO potential in terms of the bond order is particularly important for simulating the adsorption process, taking into account possible deformations \cite{Wang2009carbon,Li2014,Wang2009d} of the substrate induced by the adsorbates and \textit{viceversa}. The AIREBO potential has therefore shown good accuracy in describing adsorption of hybridized-carbon systems \cite{Petucci2013,Qi2018}. We focus on carbohydrates here for simplicity and accuracy.

\section{Results and Discussion}

\begin{figure}[htp]
\centerline{\includegraphics[width=8cm]{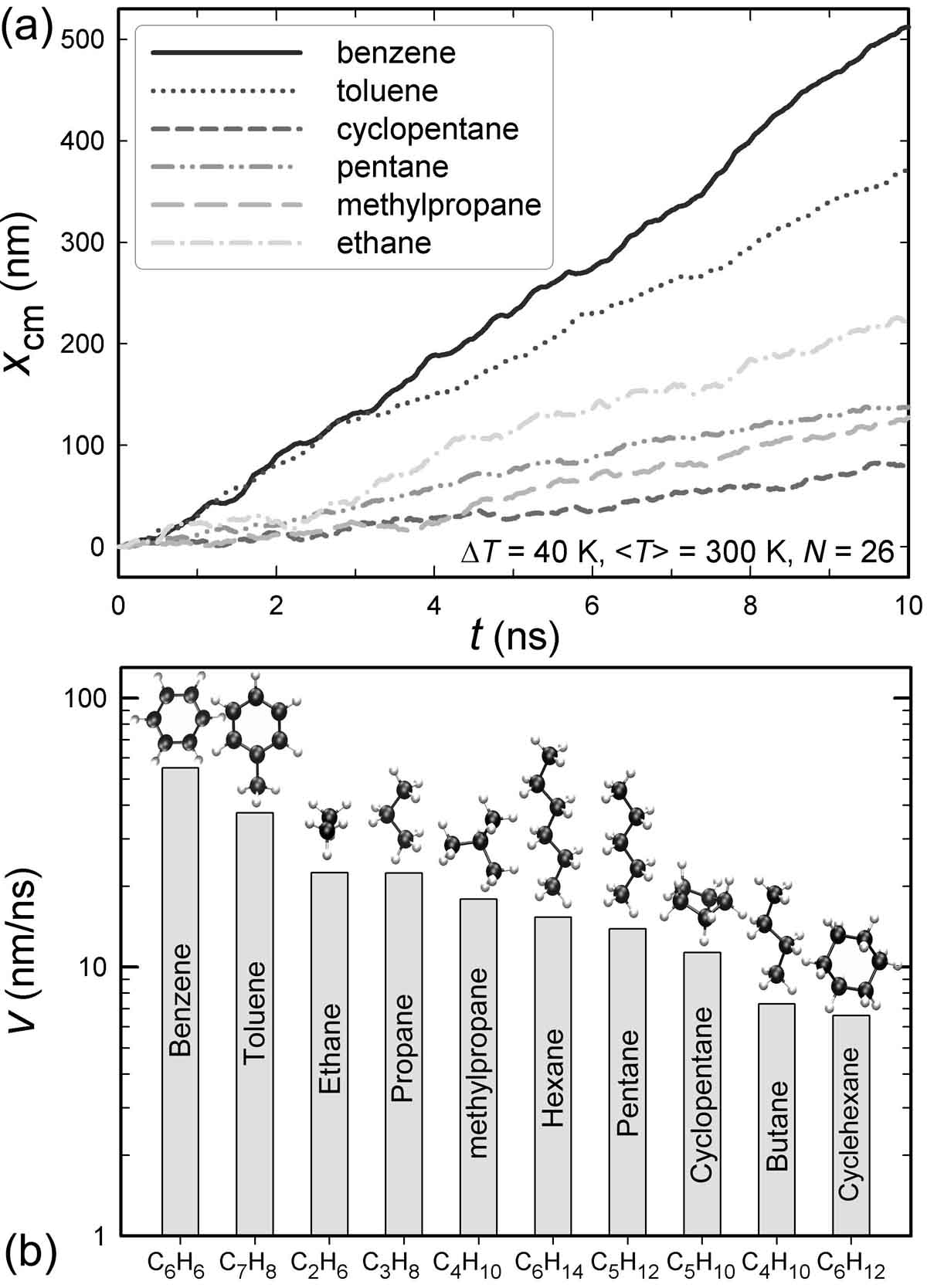}}
\caption{\label{F2}
(a) $x$ position of the center of mass of the collection of the adsorbed molecules \textit{versus} time for different molecule species. Each simulation is performed at $T_{1}=280\;\mathrm{K}$ and $T_{2}=320\;\mathrm{K}$ with $26$ adsorbates of a same type. (b) Drift velocity (in logarithmic scale) measured by taking the slope of $x_{\mathrm{cm}}$ curves at $5<t<10\;\mathrm{ns}$.}
\end{figure}

Fig.~\ref{F2}(a) shows the movement of the center of mass of molecules for different species of molecules at a constant number density. The unidirectional conduction of the molecules is found to be fast and selective. Benzene molecules are crowned champions of the race after having run more than $500\;\mathrm{nm}$ within $10 \;\mathrm{ns}$. The second place is taken by toluene molecules, which likewise have a hexagonal molecular structure. The aliphatic compounds such as ethane, pentane, methylpropane and cyclopentane move significantly slower than the aromatic ones. Video recordings in the Supporting Information illustrate the conduction of benzene and butane molecule for comparison.

Fig.~\ref{F2}(b) provides a comparison of the drift velocity between different molecule types. It can be seen that the ethane, propane, methylpropane and hexane molecules take the third, fourth, fifth and sixth positions, respectively. This can be understood from the fact that the increasing geometrical molecular cross-section increases the probability of collisions during transport. However, the subsequent ranking does not follow the expected trend of increasing molecular size. For instance, the pentane and butane molecules exhibit smaller velocity than the hexanes despite the similar structure and larger size of the latter. The cyclopentane molecules take the eighth position regardless their smaller size. The butane molecule has a similar structure to pentane and butane but smaller, and are however is even slower than cyclopentane. The last place is taken by the cyclohexane molecules, which \textit{a priori} have no big size among the tested molecules.

\begin{figure}[htp]
\centerline{\includegraphics[width=8cm]{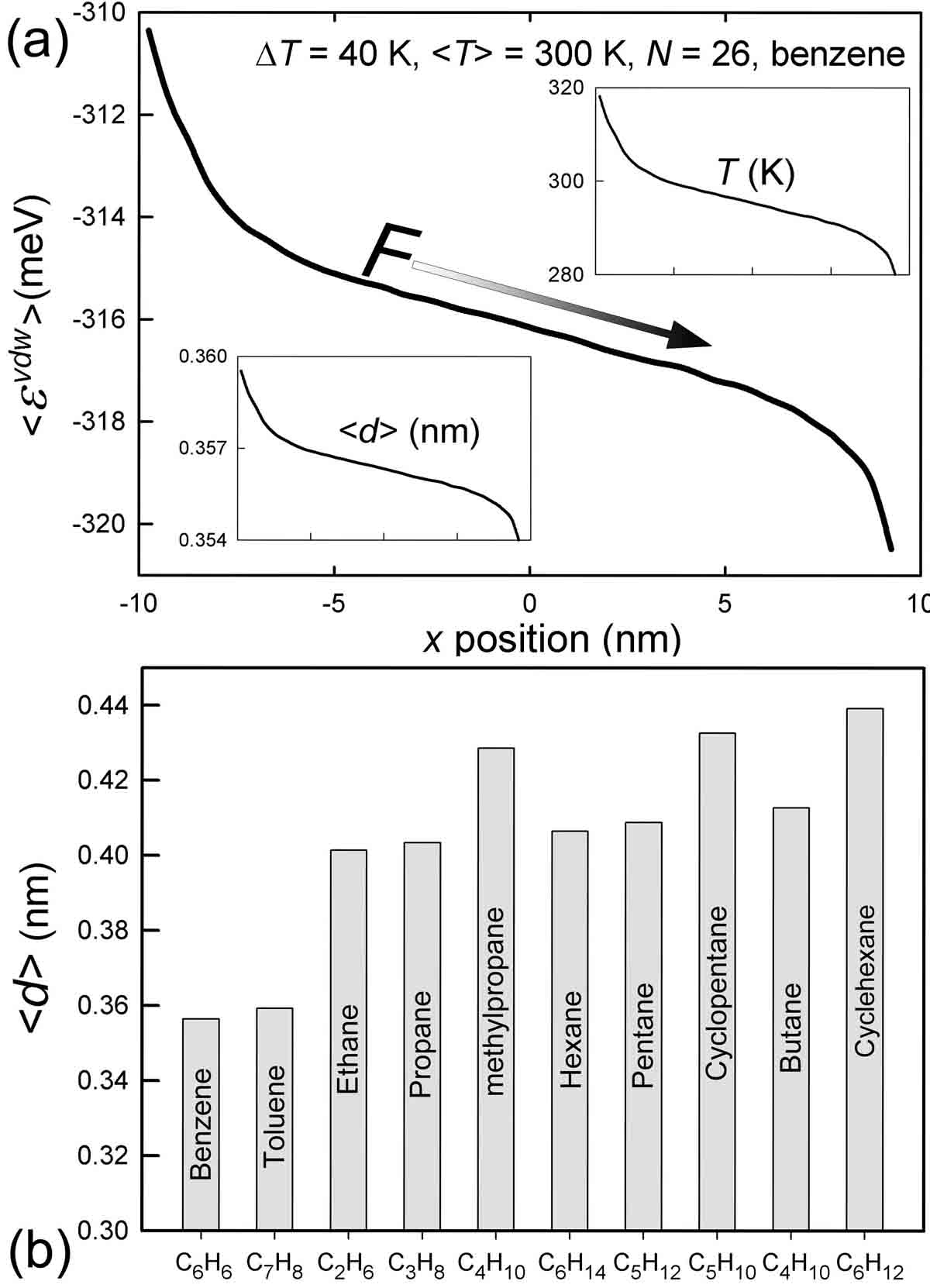}}
\caption{\label{F3}
(a) Profile of average potential energy of the interaction between the molecule and the graphene substrate. The upper inset shows the temperature gradient, the lower inset plots the time-averaged adsorbate-substrate distance $\left\langle d \right\rangle$ projected in $z$ direction. (b) $\left\langle d \right\rangle$ for different species of organic molecules.}
\end{figure}

The ranking of the molecules is striking from the perspective of textbook thermodiffusion of gaseous mixtures, in which the molecular size should be the dominating factor at a given number density of adsorbates. The selectivity for aromatic molecules clearly points to a distinct driving mechanism of molecular transport at the gas-solid interface. We find that the driving force stems from the interaction between the adsorbates and the substrate instead of from that between the adsorbates, as shown in Fig.~\ref{F3}(a). A gradient of the potential energy of the interaction between the molecules and the graphene substrate is induced by the change in their distance, due to the non-uniform temperature field shown in the insets of Fig.~\ref{F3}(a). It is seen that the temperature gradient is steeper in the regions close to the thermal reservoirs. Acceleration of the molecules can be expected in these regions due to higher driving force, despite the kinetic energy is supposed to distribute uniformly inside the reservoirs, since the size of the reservoir ($0.6\;\mathrm{nm}$ in length) is quite small comparing to the length of the graphene sheet ($20.2\;\mathrm{nm}$). The planar structures of the aromatic molecules help keep shorter distances to the substrate, and thus they are under higher driving forces. This is confirmed by the average adsorbate-substrate distance data plotted in Fig.~\ref{F3}(b), which is clearly in an inverse correlation with the competition results shown in Fig.~\ref{F2}(b).

\begin{figure}[htp]
\centerline{\includegraphics[width=8cm]{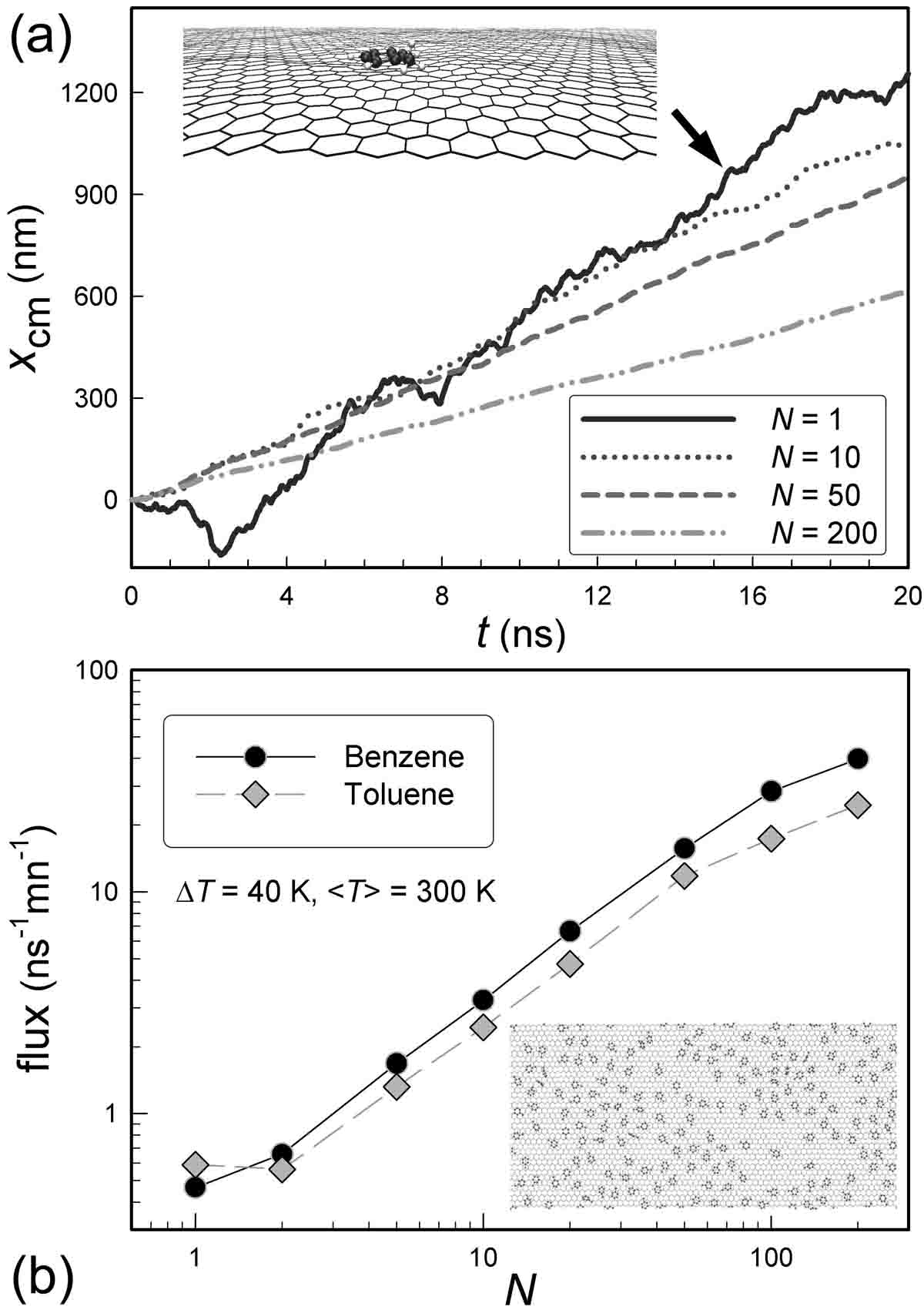}}
\caption{\label{F4}
 (a) $x_{\mathrm{cm}}$ \textit{versus} time for four simulations performed with different numbers $N$ of benzene molecules. The inset shows a snapshot of the graphene substrate at $300\;\mathrm{K}$. (b) Drift velocity \textit{versus} $N$ for benzene and toluene molecules. The inset shows a snapshot of $200$ toluene molecules drifting on graphene.}
\end{figure}

The origin of the driving force is also correlated to results of another set of simulations performed with various numbers $N$ of adsorbates, as shown in Fig.~\ref{F4}a. According to the kinetic theory of gases, the velocity should first increase and then decrease with decreasing $N$ in a gaseous mixture. However, this is not observed here. Instead, the molecules travel longer distance at lower density. A single adsorbate run with a high speed of about $60\;\mathrm{m/s}$. The increase of the drift velocities with decreasing $N$ at low molecular densities clearly indicates that the molecular transport is mainly driven by the adsorbate-graphene interaction. Another dissimilarity in the thermodiffusion at gas-solid interface and that in a gas mixture can be seen in the $N=1$ curve in Fig.~\ref{F4}a. Because of no collision with other adsorbates, a single molecule can be expected to exhibit ballistic transport, by which its $x_{cm}$ curve should have a parabolic shape. However, this is neither observed. The diffusive manner of single-molecule transport stems from thermal effects including the random rotation of the molecule and also possible collisions with thermally induced waves and ripples in the CNT surface as shown by the inset of Fig.~\ref{F4}a \cite{Fasolino2007,Barreiro2008,Wang2011,Guo2013}. This rather stands in contrast to the previously reported ballistic thermophoresis of gold on graphene \cite{Panizon2017}.

To have a comprehensive view of the density influence of the molecular transport, we plot the diffusion flux in Fig.\ref{F4}b as the amount of substance across a give $y$-direction line in the 2D graphene plane. It can be seen that the flux roughly hold a linear relation with $N$, except for too small ($N=1$) and too large density ($N>100$). This observation identifies at least two different regimes of molecular transport, which is correlated to the surface coverage. We note that the nonlinear effect for $N=1$ may also be due to the reduced sample population to make valid statistics as we see in Fig.~\ref{F4} that oscillation of $x_{cm}$ curves is more significant for small $N$. Moreover, we see that the flux of benzene molecules is kept higher than that of toluene ones at almost all density range. 

\begin{figure}[htp]
\centerline{\includegraphics[width=8cm]{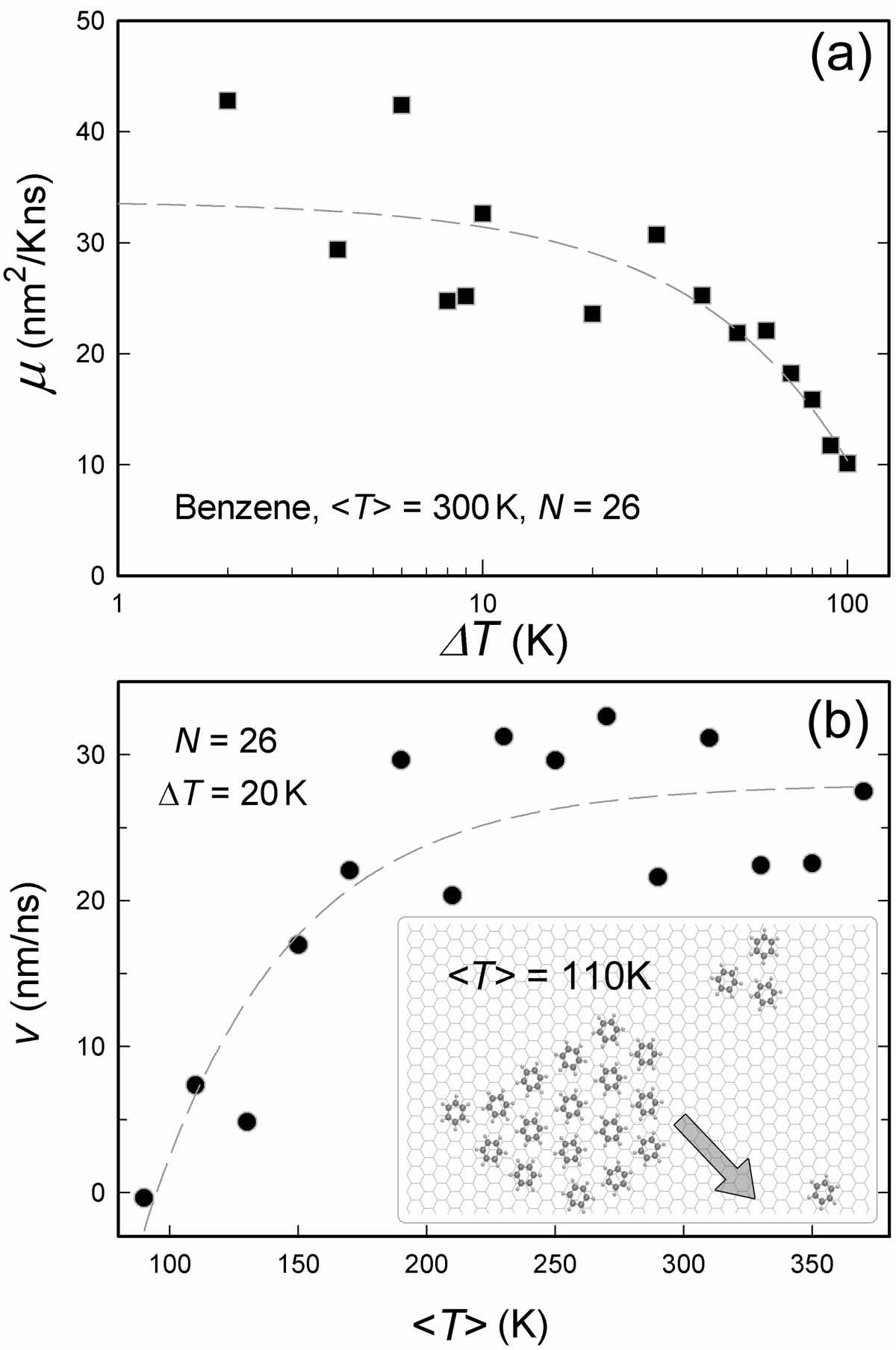}}
\caption{\label{F5}
(a) Drift mobility \textit{versus} temperature difference $\Delta T=T_{1}-T_{2}$ for 26 benzene molecules on graphene at average temperature $<T>=(T_{1}+T_{2})/2=300\;\mathrm{K}$. (b) Drift velocity at different $<T>$ with a constant $\Delta T=T_{1}-T_{2}=20\;\mathrm{K}$. The inset shows molecular aggregates on graphene formed at $110\;\mathrm{K}$.}
\end{figure}

Drift mobility is defined as $\mu=-v/{\nabla}T$ where ${\nabla}T=-{\Delta}T/L$ is the temperature gradient. $\mu$ is usually a constant for a given molecule density. However, a strong non-linear dependence of the drift velocity on the temperature gradient is observed for large ${\Delta}T$ as shown in Fig.~\ref{F5}(a). It is seen that $\mu$ decreases rapidly with increasing ${\Delta}T$ when ${\Delta}T>30\;\mathrm{K}$ (corresponding to ${\nabla}T \approx 1.5\;\mathrm{K/nm}$). This indicates a non-linear regime of the drift velocity. Two possible origins of this non-linear regime might be correlated with the scattering of molecules with flexural phonon waves in the graphene substrate \cite{Guo2013,Panizon2017}, as well as with the non-linearity of the temperature profile shown in the upper inset of Fig.~\ref{F3}(a), which becomes more significant with increasing ${\Delta}T$ \cite{Osman2001,Yang2015}. Note that $\mu$ is expected to change with the molecular type and density, as well as with the average temperature $\left\langle T \right\rangle=\left(T_{1}+T_{2}\right)/2$. Moreover, the oscillation of $\mu$ at small ${\Delta}T$ is caused by the fact that a far longer simulation time is needed to make valid statistics of the results for small ${\Delta}T$. Moreover, dissociation of a few molecules is observed at very high molecular density as shown in a video recording in the Supporting Information.

Fig.~\ref{F5}(b) shows temperature effects on the drift velocity. No conduction is observable when $\left\langle T_{a} \right\rangle$ is below an activation temperature of $100\;\mathrm{K}$. The increase of $v$ at increasing temperature is due to combined temperature effects on both the molecular diffusivity and the driving force. The diffusion coefficient of gases is known to increase at increasing temperature. Meanwhile, the driving force is supposed to be enhanced and then diminished since it is strongly correlated with the slope of the potential energy curves in relation to the adsorbate-substrate distance as the molecules can be considered as anharmonic oscillators. This could the reason why $v$ increases at increasing $\left\langle T \right\rangle$ with decreasing proportionality. Moreover, the benzenes form aggregates and move collectively at low temperature as shown in the inset of Fig.~\ref{F5}(b) as well as in a video recording in the Supporting Information. Their movement shows preference on certain orientations on the graphene surface. This is consistent with the observation by Schoen \textit{et al.} about the motion of a gold nanoparticle in helical tracks on a CNT surface \cite{Schoen2006,Mao2002} based on the registry-dependent interaction between $sp^{2}$ hybridized carbons \cite{Chen2013}.

\section{Conclusions}
The first race of organic molecules is made on a free-standing graphene substrate in MD simulations. The winners are two aromatics, namely benzene and toluene, which are found to exhibit higher drift velocities than the aliphatic molecules. This selective molecular transport is driven by a potential gradient coming from varying adsorbate-substrate distance caused by the non-uniform temperature field. Such a mechanism is quite different from that behind conventional gas-phase thermodiffusion. The molecular flux increases at increasing molecular density in two different regimes that are correlated to surface coverage. No ballistic transport is observable even for a single molecule on the graphene, probably due to thermal rippling of graphene. Furthermore, the drift mobility roughly remains constant except for under very large thermal gradient, signifying a nonlinear regime of the drift velocity. A critical activation temperature for the transport of benzenes on graphene is also detected. These results have strong implications for our understanding of the conduction of organic molecules on 2D nanostructures, and pave a way to a practical mean for selectively recognizing planar aromatic molecules, benzene in particular.


\end{document}